\UseRawInputEncoding

\documentclass[%
 aip,
 jmp,%
 amsmath,amssymb,
reprint,%
]{revtex4-1}

\usepackage{graphicx}
\usepackage{dcolumn}
\usepackage{bm}

\begin{document}


\title[Some spin textures relevant to magnetic moments in the octonion spaces]{Some spin textures relevant to magnetic moments in the octonion spaces}

\author{Zi-Hua Weng}
 \email{xmuwzh@xmu.edu.cn.}
\affiliation{
School of Aerospace Engineering, Xiamen University, Xiamen 361005, China\\
College of Physical Science and Technology, Xiamen University, Xiamen 361005, China
}%


\date{\today}

\begin{abstract}
The paper aims to apply the octonions to explore the contribution of some influence factors to magnetic moments, revealing the connection among the influence factors and spin texture. J. C. Maxwell was the first to introduce the quaternions to describe the electromagnetic theory. The subsequent scholars utilize the quaternions and octonions to study the gravitational and electromagnetic theories and so forth. The paper adopts the octonions to research the gravitational and electromagnetic fields, including the octonion angular momentum, torque, and force and so on. When the octonion force is equal to zero under some circumstances, it is able to achieve eight equations independent of each other. In particular, one of eight independent equations reveals the interrelation between the second-torque and the divergence of magnetic moments. One of its deductions is that the directions, magnitudes and frequencies of some terms are capable of impacting the orientation and vortex of magnetic moments, altering the frequency of magnetic vortex clusters. It means that the spin textures are relevant to some external influence factors. Some terms may have an influence on the arrangements of spin textures. The study will be helpful for understanding the physical properties of magnetic skyrmions and merons.
\end{abstract}


\maketitle


\section{\label{sec:level1}Introduction}

Is there any vortex associated merely with the rotating electric currents, which is independent of the vortices within ordinary fluids? Are the magnetic vortex clusters, within rotating electric currents, similar to the Karman vortex streets under some circumstances? Can the arrangement patterns of spin textures be affected by some external influence factors, such as the electromagnetic strengths and magnetic moments? For a long time, these fascinating problems attract and perplex scholars. It was not until the emergence of gravitational and electromagnetic theories, described by the octonions, that these puzzles were answered to a certain extent. According to the field theory, the periodic second-torques (in Section 2) will cause the magnetic moments to alter periodically, in the rotating electric currents. Transferring the arrangement patterns of second-torques can vary the magnetic moments and spin textures. These researches will help to understand the physical properties of magnetic skyrmions and merons.

In the chiral magnets and frustration magnets, the magnetic moments of spins are able to constitute various structural patterns of spin textures \cite{ams3,ams4}. In most cases, the spins of adjacent atoms tend to be parallel or antiparallel to each other, resulting in the ferromagnets or antiferromagnets. However, there are a few unusual physical interactions among several spins, due to the special crystals or multilayered structures. So the spins are arranged in a complicated way in some chiral magnets \cite{ams5,ams6}. Apparently, the rotating magnetic skyrmion lattices in the frustrated magnets may be more complex and diverse than those in the chiral magnets.

The magnetic skyrmion is a typical spin texture, and a small magnetic whirl. In the magnetic skyrmions, the spin directions gradually rotate from the upward direction of texture edge to the downward direction of core region, and vice versa \cite{ms1,ms2}. In the chiral magnets, the merons and antimerons have different topological properties from skyrmions. In the merons and antimerons, the spin directions in the core region of spin textures orient up or down, but the spin directions of texture edge are in the horizontal plane.

The skyrmion is a topological soliton. It is a nontrivial classical solution obtained by T. Skyrme in 1962, when he attempted to solve the nonlinear sigma model in the high energy physics. Nowadays it often refers to the skyrmions in the magnetic materials. In the asymmetric chiral magnets, it is able to form a stable ground state of the magnetic skyrmions. The arrangement patterns of magnetic skyrmions, within the chiral magnets and frustration magnets and others, are similar to those of the Karman vortex streets in the ordinary fluids, including the bi-skyrmions \cite{bis1,bis2} and alternative skyrmions and others.

The skyrmion possesses a miniature magnetic field. The spins point in different directions inside the skyrmion particles, so they are difficult to stick together for the spin magnetic fields, when the spins attempt to get close to each other. The skyrmion is a type of spin texture on the micro/nano scale. The magnetic skyrmion has a high stability due to the topology protection, which can be driven by a quite tiny electric current. Consequently, as one type of information carrier in the future memory devices, the magnetic skyrmions are widely considered to have the characteristics of high speed, high density, and low energy consumption and so forth.

J. C. Maxwell firstly utilized the algebra of quaternions to study the physical properties of electromagnetic fields. This approach inspired the subsequent scholars to utilize the quaternions and octonions to analyze the electromagnetic theory \cite{mironov,gogberashvili}, gravitational theory, hydrodynamics \cite{tanisli}, rotating electric currents \cite{chanyal}, electro-fluids, plasmas \cite{demir}, vortices of electric currents, quantum mechanics \cite{bernevig,deleo}, relativity theories \cite{manogue,bossard}, gauge theory \cite{majid,farrill}, strong nuclear force \cite{furey,bisht}, weak nuclear force, magnetic monopole \cite{add1,add2}, and condensed matter \cite{add3,add4} and so on.

The application of octonions is capable of describing simultaneously the physical quantities of electromagnetic and gravitational fields, such as, octonion angular momentum, torque, and force. When the octonion force (in Section 2) is equal to zero under some circumstances, it is able to generate eight equations independent of each other. One of eight equations reveals that the second-torques can make a contribution to the magnetic moments of the electric currents to a certain extent, including the orientation, divergence, and frequency and others.

1) Second-torque. In the rotating electric currents, the crystal defects (or dopant and others) can be inserted laterally into the electric currents. By means of the physical properties of electric currents, the crystal defects seized of the second-torques (in Section 4), within the chiral magnets and frustration magnets, will be able to transfer the second-torques into the surrounding electric currents, impacting the flow orientations of electric currents. Apparently, there is a direct relationship between the second-torque and the velocity circulation of rotating electric currents. In other words, the second-torque has a direct contribution to the movement directions of electric currents.

2) Divergence of magnetic moments. The second-torque is capable of impacting the magnetic moments of the vortices of rotating electric currents. The crystal defects may transfer the second-torques into the rotating electric currents, transforming the magnetic moments within the chiral magnets and frustration magnets, including the magnitude, curl, and divergence. Making use of the variations of second-torques, one can impact the curl and divergence of magnetic moments of rotating electric currents, shifting the orientations of rotating electric currents. Obviously, there is a relationship between the second-torque and the divergence of magnetic moments.

3) Magnetic vortex clusters. According to the second-torque continuity equation (in Section 4), if the second-torque is one periodic function, the induced magnetic moments and magnetic vortex clusters \cite{mvc1,mvc2} of rotating electric currents must also be periodic functions. Similarly, the relevant velocity circulation within the electric currents will be shifted periodically too. In terms of the electric currents composed of some types of charged particles, the variations of frequencies of the periodic second-torques will simultaneously transform the frequencies of magnetic moments and angular momenta.

By means of the electromagnetic and gravitational theories described with the octonions, the paper explores the relationship between the second-torque and the divergence of magnetic moments of rotating electric currents, discussing a few influence factors of the magnetic vortex clusters \cite{mvc3} and spin textures within the rotating electric currents (Table 1).

\begin{table}[h]
\caption{Comparison of the physical properties of vortices between the electric current and ordinary fluid.}\label{tab1}%
\begin{ruledtabular}
\begin{tabular}{lll}
subspace                       &   2-quaternion space $\mathbb{H}_e$                   &    quaternion space $\mathbb{H}_g$            \\
\hline
fluid                          &   various electric currents                           &    ordinary fluids                            \\
vortex streets                 &   magnetic vortex clusters                            &    Karman vortex streets                      \\
vorticity                      &   curl of velocity in $\mathbb{H}_e$                  &    curl of velocity in $\mathbb{H}_g$         \\
vortex                         &   rotating electric currents                          &    rotating fluids                            \\
curl                           &   curl of magnetic moments                            &    curl of linear momenta                     \\
\end{tabular}
\end{ruledtabular}
\end{table}

\section{Octonion angular momentum}

R. Descartes perceived that `the space is the extension of substance'. At present, the Cartesian academic thought is improved to `the fundamental space is the extension of fundamental field'. The fundamental field incorporates the gravitational field and electromagnetic field. Any fundamental field possesses a fundamental space. Each fundamental space is chosen as the quaternion space. Two independent quaternion spaces can be combined together to become one octonion space.

In the gravitational and electromagnetic fields without considering the contribution of material media, making use of the algebra of octonions, it is able to define the octonion field potential, field strength, field source, and linear momentum. Further, it is capable of achieving the octonion angular momentum, torque, and force of the electromagnetic and gravitational fields, including the magnetic moment, electric moment, energy, torque, angular momentum, power, and force and so forth (Table 2).

\subsection{Octonion space}

The quaternion spaces are able to describe the physical properties of gravitational fields. Also the quaternion spaces are capable of depicting the physical properties of electromagnetic fields. In the field theories, the space is regarded as the extension of one fundamental field. Apparently two fundamental fields, gravitational field and electromagnetic field, are different from each other. So the quaternion space for gravitational fields is independent of that for electromagnetic fields. In order to distinguish two different quaternion spaces, the quaternion space $\mathbb{H}_g$ denotes the quaternion space for gravitational fields. And the 2-quaternion space $\mathbb{H}_e$ indicates the quaternion space for electromagnetic fields, while the 2-quaternion space is short for the second quaternion space.

In the paper, two independent quaternion spaces, $\mathbb{H}_g$ and $\mathbb{H}_e$ , can be considered to be perpendicular to each other. Therefore, the two quaternion spaces, $\mathbb{H}_g$ and $\mathbb{H}_e$ , can be combined together to become one octonion space $\mathbb{O}$ . In other words, the octonion space $\mathbb{O}$ is able to describe simultaneously the physical properties of gravitational and electromagnetic fields, including the octonion field potential $\mathbb{A}$ , field strength $\mathbb{F}$ , field source $\mathbb{S}$ , linear momentum $\mathbb{P}$, angular momentum $\mathbb{L}$ , torque $\mathbb{W}$ , and force $\mathbb{N}$ and so forth \cite{weng1,weng2}.

In the quaternion space $\mathbb{H}_g$ , the radius vector is, $\mathbb{R}_g = i r_0 \textbf{i}_0 + \Sigma r_k \textbf{i}_k$ , and $\textbf{i}_j$ is the basis vector, $r_j$ is the coordinate value. The quaternion velocity is, $\mathbb{V}_g = i v_0 \textbf{i}_0 + \Sigma v_k \textbf{i}_k$. In the 2-quaternion space $\mathbb{H}_e$ , the radius vector is, $\mathbb{R}_e = i R_0 \textbf{I}_0 + \Sigma R_k \textbf{I}_k$ , and $\textbf{I}_j$ is the basis vector, $R_j$ is the coordinate value. The 2-quaternion velocity is, $\mathbb{V}_e = i V_0 \textbf{I}_0 + \Sigma V_k \textbf{I}_k$ . Consequently, in the octonion space $\mathbb{O}$ , the octonion radius vector is, $\mathbb{R} = \mathbb{R}_g + k_{eg} \mathbb{R}_e$ , and the octonion velocity is, $\mathbb{V} = \mathbb{V}_g + k_{eg} \mathbb{V}_e$ .
Herein $k_{eg}$ is one coefficient, to ensure the dimensional homogeneity. $r_j$ , $R_j$ , $v_j$ and $V_j$ are all real. $i$ is the imaginary unit. The symbol $\circ$ denotes the octonion multiplication. $\textbf{i}_0 = 1$. $\textbf{i}_k \circ \textbf{i}_k = -1$. $\textbf{I}_j \circ \textbf{I}_j = -1$. $\textbf{I}_k = \textbf{i}_k \circ \textbf{I}_0 $. $k = 1, 2, 3$. $j = 0, 1, 2, 3$.

\subsection{Angular momentum}

From the octonion linear momentum $\mathbb{P}$ and radius vector $\mathbb{R}$ and others (see Ref.[29]), it is able to define the octonion angular momentum, $\mathbb{L} = (\mathbb{R} + k_{rx} \mathbb{X} )^\times \circ \mathbb{P}$ . Where $\mathbb{P} = \mathbb{P}_g + k_{eg} \mathbb{P}_e$ . $\mathbb{P}_g$ is the component of the octonion linear momentum $\mathbb{P}$ in the quaternion space $\mathbb{H}_g$ , while $\mathbb{P}_e$ is that in the 2-quaternion space $\mathbb{H}_e$ . $\mathbb{X} = \mathbb{X}_g + k_{eg} \mathbb{X}_e$ . $\mathbb{X}_g$ is the constituent of the octonion integrating function $\mathbb{X}$ of field potential in the quaternion space $\mathbb{H}_g$ , while $\mathbb{X}_e$ is that in the 2-quaternion space $\mathbb{H}_e$ . $\ast$ and $\times$ denote the octonion conjugate and complex conjugate, respectively. $k_{rx}$ is the coefficient, to ensure the dimensional homogeneity.

In the above, $\mathbb{L} = \mathbb{L}_g + k_{eg} \mathbb{L}_e$ . $\mathbb{L}_g$ is the ingredient of the octonion angular momentum $\mathbb{L}$ in the quaternion space $\mathbb{H}_g$ , while $\mathbb{L}_e$ is that in the 2-quaternion space $\mathbb{H}_e$. $\mathbb{L}_g = L_{10} + i \textbf{L}_1^i + \textbf{L}_1$ . $\mathbb{L}_e = \textbf{L}_{20} + i \textbf{L}_2^i + \textbf{L}_2$. $\textbf{L}_1$ is the angular momentum. $\textbf{L}_2^i$ is the electric moment, and $\textbf{L}_2$ is the magnetic moment. Herein $\textbf{L}_1 = \Sigma L_{1k} \textbf{i}_k$ , $\textbf{L}_1^i = \Sigma L_{1k}^i \textbf{i}_k$. $\textbf{L}_{20} = L_{20} \textbf{I}_0$ . $\textbf{L}_2 = \Sigma L_{2k} \textbf{I}_k$ , $\textbf{L}_2^i = \Sigma L_{2k}^i \textbf{I}_k$. $L_{1j}$, $L_{1k}^i$ , $L_{2j}$ , and $L_{2k}^i$ are all real.

\begin{table}[h]
\caption{Some physical quantities of gravitational and electromagnetic fields, in the octonion space $\mathbb{O}$ without considering the contribution of material media.}\label{tab2}%
\begin{ruledtabular}
\begin{tabular}{ll}
physical quantity                      &   definition                                                                                                          \\
\hline
octonion field source                  &   $\mu \mathbb{S} = - ( i \mathbb{F} / v_0 + \lozenge )^\ast \circ \mathbb{F}$                                        \\
octonion linear momentum               &   $\mathbb{P} = \mu \mathbb{S} / \mu_g$                                                                               \\
octonion angular momentum              &   $\mathbb{L} = ( \mathbb{R} + k_{rx} \mathbb{X} )^\times \circ \mathbb{P} $                                          \\
octonion torque                        &   $\mathbb{W} = - v_0 ( i \mathbb{F} / v_0 + \lozenge ) \circ \{ ( i \mathbb{V}^\times / v_0  ) \circ \mathbb{L} \}$  \\
octonion force                         &   $\mathbb{N} = - ( i \mathbb{F} / v_0 + \lozenge ) \circ \{ ( i \mathbb{V}^\times / v_0  ) \circ \mathbb{W} \}$      \\
\end{tabular}
\end{ruledtabular}
\end{table}

\section{Octonion composite force}

In the octonion space $\mathbb{O}$ , the octonion field strength $\mathbb{F}$ and angular momentum $\mathbb{L}$ can be combined together to become one octonion composite field strength, $\mathbb{F}^+ = \mathbb{F} + k_{fl} \mathbb{L}$ . $k_{fl}$ is a coefficient, to meet the need of dimensional homogeneity. Compared with the classical electromagnetic theory, it can be found that the new physical quantity $\mathbb{F}^+$ is the field strength within the material media (Table 3).

From the octonion composite field strength $\mathbb{F}^+$ , it is able to achieve successionally the octonion composite field source $\mathbb{S}^+$, composite linear momentum $\mathbb{P}^+$ , composite angular momentum $\mathbb{L}^+$ , composite torque $\mathbb{W}^+$ , and composite force $\mathbb{N}^+$ and so forth  within the material media. And the octonion composite property equation is, $\mathbb{P}^+ = k_{pf} \mathbb{F}^+$ , with $k_{pf}$ being one coefficient relevant to the electromagnetic and gravitational properties (see Ref.[29]).

\subsection{Composite torque}

From the octonion composite angular momentum $\mathbb{L}^+$ and composite field strength $\mathbb{F}^+$ and others, it is able to define the octonion composite torque, $\mathbb{W}^+ =  - v_0 ( i \mathbb{F}^+ / v_0 + \lozenge ) \circ \{ ( i \mathbb{V}^\times / v_0  ) \circ \mathbb{L}^+  \}$ , within the material media. Where $\mathbb{F}^+ = \mathbb{F}_g^+ + k_{eg} \mathbb{F}_e^+$ . The gravitational strength $\mathbb{F}_g^+$ is the component of the octonion composite field strength $\mathbb{F}^+$ in the quaternion space $\mathbb{H}_g$, while the electromagnetic strength $\mathbb{F}_e^+$ is that in the 2-quaternion space $\mathbb{H}_e$ . The gravitational strength $\mathbb{F}_g^+$ consists of the gravitational acceleration $\textbf{g}^+$ , and gravitational precessional-angular-velocity $\textbf{b}^+$  within the gravitational media \cite{weng3,weng4}. The electromagnetic strength $\mathbb{F}_e^+$ comprise the electric field intensity $\textbf{E}^+$, and magnetic flux density $\textbf{B}^+$  within the electromagnetic media. $\lozenge$ is the quaternion operator. $\lozenge = i \partial_0 + \nabla$ . $\nabla = \Sigma \textbf{i}_k \partial_k$, $\partial_j = \partial / \partial r_j$ . $r_0 = v_0 t$ . $v_0$ is the speed of light, while $t$ is the time.

In the above, $\mathbb{W}^+ = \mathbb{W}_g^+ + k_{eg} \mathbb{W}_e^+$ . $\mathbb{W}_g^+$ is the ingredient of the octonion composite torque $\mathbb{W}^+$ in the quaternion space $\mathbb{H}_g$, while $\mathbb{W}_e^+$ is that in the 2-quaternion space $\mathbb{H}_e$ . $\mathbb{W}_g^+ = i W_{10}^{i+} + W_{10}^+ + i \textbf{W}_1^{i+} + \textbf{W}_1^+$. $\mathbb{W}_e^+ = i \textbf{W}_{20}^{i+} + \textbf{W}_{20}^+ + i \textbf{W}_2^{i+} + \textbf{W}_2^+$. $W_{10}^{i+}$ is the energy within the material media. $\textbf{W}_1^{i+}$ is the torque within the material media, including the gyroscopic torque, $i \nabla ( \textbf{v} \cdot \textbf{L}_1^+ )$. $W_{10}^+$ is the divergence of angular momentum $\textbf{L}_1^+$ . $\textbf{W}_1^+$ is the curl of angular momentum $\textbf{L}_1^+$. $\textbf{W}_{20}^{i+}$ is the second-energy within the material media, and $\textbf{W}_2^{i+}$ is the second-torque within the material media. $\textbf{W}_{20}^+$ is the divergence of magnetic moment $\textbf{L}_2^+$ . $\textbf{W}_2^+$ is the curl of magnetic moment $\textbf{L}_2^+$ . Herein $\textbf{W}_1^+ = \Sigma W_{1k}^+ \textbf{i}_k$ , $\textbf{W}_1^{i+} = \Sigma W_{1k}^{i+} \textbf{i}_k$. $\textbf{W}_{20}^+ = W_{20}^+ \textbf{I}_0$ . $\textbf{W}_{20}^{i+} = W_{20}^{i+} \textbf{I}_0$. $\textbf{W}_2^+ = \Sigma W_{2k}^+ \textbf{I}_k$ , $\textbf{W}_2^{i+} = \Sigma W_{2k}^{i+} \textbf{I}_k$ . $W_{1j}^+$ , $W_{1j}^{i+}$ , $W_{2j}^+$ , and $W_{2j}^{i+}$ are all real.

\subsection{Composite force}

Furthermore, from the octonion composite torque $\mathbb{W}^+$ and composite field strength $\mathbb{F}^+$ and others, it is able to define the octonion composite force, $\mathbb{N}^+ = - ( i \mathbb{F}^+ / v_0 + \lozenge ) \circ \{ ( i \mathbb{V}^\times / v_0  ) \circ \mathbb{W}^+  \}$ , within the material media. Where $\mathbb{N}^+ = \mathbb{N}_g^+ + k_{eg} \mathbb{N}_e^+$ . $\mathbb{N}_g^+$ is the constituent of the octonion composite force $\mathbb{N}^+$ in the quaternion space $\mathbb{H}_g$ , while $\mathbb{N}_e^+$ is that in the 2-quaternion space $\mathbb{H}_e$ . $\mathbb{N}_g^+ = i N_{10}^{i+} + N_{10}^+ + i \textbf{N}_1^{i+} + \textbf{N}_1^+$. $\mathbb{N}_e^+ = i \textbf{N}_{20}^{i+} + \textbf{N}_{20}^+ + i \textbf{N}_2^{i+} + \textbf{N}_2^+$. $N_{10}^{i+}$ is the divergence of torque $\textbf{W}_1^{i+}$ , and relevant to the torque continuity equation within the material media. $\textbf{N}_1^{i+}$ is the force, including the Magnus force, $ \nabla ( v_0 \partial_0 L_{10}^+ )$, within the material media. And it is related with the force equilibrium equation within the material media. $N_{10}^+$ is the power within the material media, which is connected with the fluid continuity equation within the material media. $\textbf{N}_1^+$ is the derivative of torque $\textbf{W}_1^{i+}$, and interrelated with the precession equilibrium equation within the material media. $\textbf{N}_{20}^+$ is the second-power, and bound up with the current continuity equation within the material media. $\textbf{N}_1^+ = \Sigma N_{1k}^+ \textbf{i}_k$ , $\textbf{N}_1^{i+} = \Sigma N_{1k}^{i+} \textbf{i}_k$. $\textbf{N}_{20}^+ = N_{20}^+ \textbf{I}_0$ . $\textbf{N}_{20}^{i+} = N_{20}^{i+} \textbf{I}_0$. $\textbf{N}_2^+ = \Sigma N_{2k}^+ \textbf{I}_k$ , $\textbf{N}_2^{i+} = \Sigma N_{2k}^{i+} \textbf{I}_k$ . $N_{1j}^+$ , $N_{1j}^{i+}$ , $N_{2j}^+$ , and $N_{2j}^{i+}$ are all real.

In certain circumstances, the octonion composite force may be equal to zero. As a result, it is able to attain eight independent equations from $\mathbb{N}^+ = 0$, achieving the continuity equations or equilibrium equations, in the gravitational and electromagnetic fields considering the contribution of material media.

\begin{table*}
\caption{Several composite physical quantities of gravitational and electromagnetic fields, in the octonion space $\mathbb{O}$ considering the contribution of material media.}\label{tab3}%
\begin{ruledtabular}
\begin{tabular}{ll}
composite physical quantity            &   formula                                                                                                                 \\
\hline
octonion composite field strength      &   $\mathbb{F}^+ = \mathbb{F} + k_{fl} \mathbb{L}$                                                                         \\
octonion composite field source        &   $\mu \mathbb{S}^+ = - ( i \mathbb{F}^+ / v_0 + \lozenge )^\ast \circ \mathbb{F}^+$                                      \\
octonion composite linear momentum     &   $\mathbb{P}^+ = \mu \mathbb{S}^+ / \mu_g$                                                                               \\
octonion composite property equation   &   $\mathbb{P}^+ = k_{pf} \mathbb{F}^+$                                                                                    \\
octonion composite angular momentum    &   $\mathbb{L}^+ = ( \mathbb{R} + k_{rx} \mathbb{X} )^\times \circ \mathbb{P}^+ $                                          \\
octonion composite torque              &   $\mathbb{W}^+ = - v_0 ( i \mathbb{F}^+ / v_0 + \lozenge ) \circ \{ ( i \mathbb{V}^\times / v_0 ) \circ \mathbb{L}^+ \}$ \\
octonion composite force               &   $\mathbb{N}^+ = - ( i \mathbb{F}^+ / v_0 + \lozenge ) \circ \{ ( i \mathbb{V}^\times / v_0  ) \circ \mathbb{W}^+ \}$    \\
\end{tabular}
\end{ruledtabular}
\end{table*}

\section{Second-torque continuity equation}

In the octonion space considering the contribution of material media, one can achieve the eight equilibrium equations or continuity equations for the gravitational and electromagnetic fields (Table 4), from the octonion composite force, $\mathbb{N}^+ = 0$, especially the second-torque continuity equation, $\textbf{N}_{20}^{i+} = 0$. By virtue of the latter, it is able to investigate the contribution of some external influence factors on the vortices of rotating electric currents, within the chiral magnets and frustration magnets and so forth considering the contribution of material media. The influence factors include the second-torque, gravitational strength, electromagnetic strength, and divergence of magnetic moments and so on.

The second-torque continuity equation, $\textbf{N}_{20}^{i+} = 0$, can be further expanded into (see Ref.[29]),
\begin{eqnarray}
0 = && ( \textbf{g}^+ \cdot \textbf{W}_2^{i+} / v_0 - \textbf{b}^+ \cdot \textbf{W}_2^+ ) / v_0
\\
\nonumber
&&
- ( \partial_0 \textbf{W}_{20}^+ + \nabla \cdot \textbf{W}_2^{i+} )
\\
\nonumber
&&
+ ( \textbf{E}^+ \cdot \textbf{W}_1^{i+} / v_0 - \textbf{B}^+ \cdot \textbf{W}_1^+ ) / v_0 ~.
\end{eqnarray}

The above states that the gravitational strength $( \textbf{g}^+ , \textbf{b}^+ )$, electromagnetic strength $( \textbf{E}^+ , \textbf{B}^+ )$, torque $\textbf{W}_1^{i+}$ , and curl $\textbf{W}_2^+$ of magnetic moments and others are able to make a contribution to the divergence $\textbf{W}_{20}^+$ of magnetic moment and second-torque $\textbf{W}_2^{i+}$ within material media. And even the field strength can obviously impact the divergence of magnetic moment and second-torque within material media, when the gravitational strength or electromagnetic strength are extreme strong under certain circumstances.

\subsection{Zero field strength}

When there is neither the gravitational strength nor the electromagnetic strength within material media, the second-torque continuity equation, Eq.(1), can be simplified into,
\begin{eqnarray}
- \partial ( \nabla \cdot \textbf{L}_2^+ ) / \partial t + \nabla \cdot \textbf{W}_2^{i+} = 0 ~.
\end{eqnarray}

The above states that the divergence $\textbf{W}_{20}^+$ of magnetic moments and second-torque $\textbf{W}_2^{i+}$ , within the vortices of rotating electric currents, are closely related to each other, rather than independent of each other. Further it is found that the divergence $\textbf{W}_{20}^+$ of magnetic moments is interrelated with the magnetic moments, while the second-torque $\textbf{W}_2^{i+}$ is relevant to not only the derivative of magnetic moments but also the curl of electric moments. In other words, the above reveals that the magnetic moments and electric moments are correlative with each other, within material media sometimes. And this is similar to the situation in the fluid continuity equation, in which the linear momentum and mass are interrelated with each other.

When the derivative of magnetic moments of second-torque $\textbf{W}_2^{i+}$ is equal to zero within material media, the above reveals the relationship between the divergence $\textbf{W}_{20}^+$ of magnetic moments with the curl of electric moments of second-torque $\textbf{W}_2^{i+}$. It means that the divergence $\textbf{W}_{20}^+$ of magnetic moments and the curl of electric moments are not two physical quantities independent of each other within material media.

To a certain extent, the second-torque continuity equation can be seen as an analogue to the fluid continuity equation, even in the material media. Apparently the divergence $\textbf{W}_{20}^+$ of magnetic moments and second-torque $\textbf{W}_2^{i+}$, in the second-torque continuity equation, are similar to the mass and linear momentum in the fluid continuity equation, respectively. Also it is analogous to the situation of linear momentum, the second-torque $\textbf{W}_2^{i+}$ possesses many types of variations, including the transformations of orientations and magnitudes. As a result, the divergence, $\nabla \cdot \textbf{L}_2^+$ , of magnetic moments will be seized of many types of modifications within material media.

Further, the above can be rewritten as,
\begin{eqnarray}
\nabla \cdot ( - \partial \textbf{L}_2^+ / \partial t + \textbf{W}_2^{i+} ) = 0 ~.
\end{eqnarray}

According to the second-torque continuity equation within material media, it is easy to find that the magnetic moment $\textbf{L}_2$ must be a cosine function with the phase value $\omega t$ , if the second-torque is one sinusoidal function with the phase value $\omega t$ . The frequencies of these two functions are identical. That is,
\begin{eqnarray}
L_2^+ = L_{20}^+ cos ( \omega t + \phi_1 ),~ W_2^{i+} = W_{20}^{i+} sin ( \omega t + \phi_2 ),
\end{eqnarray}
where $\textbf{L}_2^+ = \textbf{u} L_2^+$ , $\textbf{W}_2^{i+} = \textbf{u} W_2^{i+} $ . $\textbf{u}$ is an unit vector in the 2-quaternion space $\mathbb{H}_e$ . $L_{20}^+$ and $W_{20}^{i+}$ are all real. $\omega$ is the angular velocity. $\phi_1$ and $\phi_2$ both are phase angles. Apparently, the phase difference between the second-torque and magnetic moment may reach to $\pi / 2$.

\begin{table*}
\caption{In the octonion space $\mathbb{O}$ considering the contribution of material media, eight independent equations can be simultaneously derived from one single equation, $\mathbb{N}^+ = 0$, including the four equilibrium equations and four continuity equations.}\label{tab4}
\begin{ruledtabular}
\begin{tabular}{lllcccc}
equation                     &    equilibrium/continuity equation            &    application example              &     subspace           &   reference        \\
\hline
$\textbf{N}_1^+ = 0$         &    precession equilibrium equation            &    Larmor precession                &     $\mathbb{H}_g$     &   Ref.[29]         \\
$\textbf{N}_1^{i+} = 0$      &    force equilibrium equation                 &    (many)                           &     $\mathbb{H}_g$     &   --               \\
$\textbf{N}_2^+ = 0$         &    second-precession equilibrium equation     &    piezoelectric effects            &     $\mathbb{H}_e$     &   Ref.[30]         \\
$\textbf{N}_2^{i+} = 0$      &    second-force equilibrium equation          &    superconducting current,         &     $\mathbb{H}_e$     &   Ref.[31]         \\
                             &                                               &    droplet transportation           &                        &   ?                \\
$N_{10}^+ = 0$               &    fluid continuity equation                  &    (many)                           &     $\mathbb{H}_g$     &   --               \\
$N_{10}^{i+} = 0$            &    torque continuity equation                 &    Karman vortex streets            &     $\mathbb{H}_g$     &   Ref.[32]         \\
$\textbf{N}_{20}^+ = 0$      &    current continuity equation                &    (many)                           &     $\mathbb{H}_e$     &   --               \\
$\textbf{N}_{20}^{i+} = 0$   &    second-torque continuity equation          &    spin textures,                   &     $\mathbb{H}_e$     &   ?                \\
                             &                                               &    magnetic vortex clusters         &                        &   ?                \\
\end{tabular}
\end{ruledtabular}
\end{table*}

\subsection{Magnetic vortices}

The electromagnetic media may possess a few crystal defects. The rotating electric currents and crystal defects are similar to the ordinary fluids in the hydrodynamics, respectively. The flow directions of rotating electric currents can be shifted by the crystal defects. And this is analogous to the transforming the flow directions of ordinary fluids.

In the electromagnetic media, one crystal defect refers to the occurrence of structural distortion or functional abnormality at a certain place, for instance, the crystal defects and dopant and so forth in the crystals and even superconductors. In the electromagnetic media surrounding by the rotating electric currents, the effect of crystal defects is analogous to that of cylinders in the ordinary fluids. Consequently it can be compared with the research methods of Karman vortex streets, in the ordinary fluids, to explore the physical properties of the magnetic vortices (or the magnetic moments of rotating electric currents), within the electromagnetic media surrounding by the rotating electric currents (Table 5).

By comparison, it is easy to find that the arrangement patterns of magnetic vortices, within the rotating electric currents of chiral magnets and frustration magnets and so forth, are similar to those of the Karman vortex streets in the ordinary fluids. There must be also the laminar flows or turbulent flows, in terms of the flow types of various electric currents.

When the rotating electric currents flow through some crystal defects, the second-torque, $\textbf{W}_2^{i+}$ , will result in the crystal defects within the electromagnetic media to generate a few types of rotations in the 2-quaternion space $\mathbb{H}_e$ , shifting the motion directions of rotating electric currents, in the 2-quaternion space $\mathbb{H}_e$ . It means that the second-torques applied to the crystal defects can be transferred to the rotating electric currents.

In terms of the rotating electric currents, according to the second-torque continuity equation, Eq.(1), some different types of second-torques, $\textbf{W}_2^{i+}$ , will produce a variety of arrangement patterns of the magnetic vortices (Table 6), in particular the spin textures within the chiral magnets and frustration magnets and so on.

1) Symmetrical magnetic vortices. When the flow velocity of rotating electric currents is slow, the electric currents can induce the symmetrical magnetic vortices, behind a crystal defect without any rotation in the 2-quaternion space $\mathbb{H}_e$ . Two second-torques, produced by the electric currents flowing through this crystal defect, are symmetrical on both sides (see Refs.[7] and [8]). The difference of second-torques between two sides is close to zero. It can be considered as the result of two second-torques, $\textbf{W}_{2(1)}^{i+}$ and $\textbf{W}_{2(2)}^{i+}$, applied simultaneously to the crystal defect, with $\textbf{W}_{2(2)}^{i+} = - \textbf{W}_{2(1)}^{i+}$. In other words, the amplitude, frequency, and phase value of the second-torque, $\textbf{W}_{2(1)}^{i+}$, are equal to those of the second-torque, $\textbf{W}_{2(2)}^{i+}$, respectively.

2) Alternative magnetic vortices. In case we increase properly the flow velocity of electric currents, the alternately magnetic vortices may be produced behind a crystal defect with one rotation in the 2-quaternion space $\mathbb{H}_e$ , forming the magnetic vortex clusters. When the electric currents flow through either one asymmetrical crystal defect or the crystal defect with a rotation in the 2-quaternion space $\mathbb{H}_e$, two second-torques are non-symmetrical on both sides, therefore the second-torque is a periodic function.

The pressure-like difference, caused by the low-velocity electric currents on both sides of the magnetic vortex clusters, will also exert a certain squeezing effect on the alternately magnetic vortices (see Refs.[3] and [4]), forming the arrangement patterns of magnetic vortices. This is the consequence of two second-torques, $\textbf{W}_{2(3)}^{i+}$ and $\textbf{W}_{2(4)}^{i+}$, applied simultaneously to a crystal defect in the 2-quaternion space $\mathbb{H}_e$. And the amplitude and frequency of the second-torque, $\textbf{W}_{2(3)}^{i+}$, are equal to those of the second-torque, $\textbf{W}_{2(4)}^{i+}$, respectively. But the phase values of the two second-torques are different.

3) Biased magnetic vortices. When the electric currents with a low flow-velocity may flow through one asymmetric crystal defect without any rotation in the 2-quaternion space $\mathbb{H}_e$ , it also generates the magnetic vortex clusters. As a result, the second-torques are also periodic, resulting in the magnetic moments of a part of electric currents to vary periodically. It can be regarded as the result of two second-torques, $\textbf{W}_{2(5)}^{i+}$ and $\textbf{W}_{2(6)}^{i+}$, applied simultaneously to a crystal defect. The amplitudes and phase values of the two second-torques, $\textbf{W}_{2(5)}^{i+}$ and $\textbf{W}_{2(6)}^{i+}$, are different from each other, respectively, but their frequencies are the same.

4) Arbitrary magnetic vortices. Apparently an arbitrary second-torque is a type of complicated situation. It can be considered as the consequence of two second-torques, $\textbf{W}_{2(7)}^{i+}$ and $\textbf{W}_{2(8)}^{i+}$, applied simultaneously to a crystal defect, in the 2-quaternion space $\mathbb{H}_e$ . The amplitude, frequency, and phase value of the second-torque, $\textbf{W}_{2(7)}^{i+}$, may be different from those of the second-torque, $\textbf{W}_{2(8)}^{i+}$, respectively. In a word, when the second-torque applied to the rotating electric currents changes, the magnetic moments of some elements of electric currents will alter, transforming the magnetic vortex clusters accordingly.

For a few particles, the mass and electric charge can be combined together to become one charged particle. Its linear momentum locates in the quaternion space $\mathbb{H}_g$, while its electric current situates in the 2-quaternion space $\mathbb{H}_e$. In certain exceptional circumstances, the linear momentum of charged particles may be proportional to the electric current to some extent. It means that the linear momenta and magnetic moments of the rotating electric currents can be transformed simultaneously, when the second-torques are applied to the rotating electric currents.

\begin{table*}
\caption{Comparison of the physical properties between the magnetic vortex cluster in the rotating electric currents with the Karman vortex street in the ordinary fluids.}\label{tab5}%
\begin{ruledtabular}
\begin{tabular}{lll}
subspace                         &    2-quaternion space $\mathbb{H}_e$                             &    quaternion space $\mathbb{H}_g$                     \\
\hline
miniature-object                 &    crystal defects and dopant within electric currents           &    cylinder and others within the ordinary fluids      \\
vortex                           &    vortices of rotating electric currents	                    &    vortices of ordinary fluids                         \\
spatial dimension                &    2-D, or 3-D magnetic vortex clusters                          &    2-D Karman vortex streets, or 3-D vortex rings      \\
transfer                         &    from second-torques to electric currents                      &    from torques to ordinary fluids                     \\
\end{tabular}
\end{ruledtabular}
\end{table*}

\subsection{Zero second-torque}

When the divergence of second-torque is equal to zero, that is, $\nabla \cdot \textbf{W}_2^{i+} = 0$, the second-torque continuity equation, Eq.(2), within material media can be degenerated into,
\begin{eqnarray}
\nabla \cdot ( \partial \textbf{L}_2^+ / \partial t ) = 0 ~,
\end{eqnarray}
one of solutions is, $\partial \textbf{L}_2^+ / \partial t = 0$ . On this condition, the magnetic moment $\textbf{L}_2^+$ of the rotating electric currents must be a vector quantity that does not change with time, within some chiral magnets and frustration magnets.

In the second-torque continuity equation within material media, two operators, $\nabla$ and $\partial / \partial t$ , can be considered to be commutative with each other. As a result, the above can be rewritten as,
\begin{eqnarray}
\partial ( \nabla \cdot \textbf{L}_2^+ ) / \partial t = 0 ~,
\end{eqnarray}
one of solutions is, $\nabla \cdot \textbf{L}_2^+ = scalar$. In this instance, the divergence, $\nabla \cdot \textbf{L}_2^+$ , of magnetic moments of the rotating electric currents will be a scalar quantity that does not change with time, within the chiral magnets and frustration magnets and so on.

\subsection{Field strength}

According to Eq.(1), it is easy to find that the field strength and second-torque should match each other in the electromagnetic media, in order to play an important role in the rotating electric currents.

Eq.(1) can be rewritten as,
\begin{eqnarray}
\textbf{f}_c + ( \partial_0 \textbf{W}_{20}^+ + \nabla \cdot \textbf{W}_2^{i+} ) = 0 ~,
\end{eqnarray}
where $- \textbf{f}_c = ( \textbf{g}^+ \cdot \textbf{W}_2^{i+} / v_0 - \textbf{b}^+ \cdot \textbf{W}_2^+ ) / v_0 + ( \textbf{E}^+ \cdot \textbf{W}_1^{i+} / v_0 - \textbf{B}^+ \cdot \textbf{W}_1^+ ) / v_0 $ .

a) When $\textbf{f}_c + \partial_0 \textbf{W}_{20}^+ = 0$ , there is $\nabla \cdot \textbf{W}_2^{i+} = 0$ . Its solution is similar to that of Eq.(3). It means that there may be a few solutions for $\nabla \cdot \textbf{W}_2^{i+} = 0$, including $\textbf{W}_2^{i+} = 0$.

b) When $\textbf{f}_c + \partial_0 \textbf{W}_{20}^+ \neq 0 $ , Eq.(7) has some solutions similar to those of Eq.(2). One of these solutions is, $\textbf{f}_c + \partial_0 \textbf{W}_{20}^+ = - \partial ( \nabla \cdot \textbf{L}_2' ) / \partial t$ , with $\textbf{L}_2'$ being the equivalent magnetic moment in the 2-quaternion space $\mathbb{H}_e$ . This is one magnetic vortex cluster under certain octonion field strengths. However the magnetic vortex clusters, under certain octonion field strengths, must be different to those without any octonion field strength.

Apparently the octonion field strength has a certain contribution to the magnetic vortex clusters. In the definition of $\mathbb{W}^+$, the octonion field strength will contribute to the terms of octonion torque, in particular the second-torque $\textbf{W}_2^{i+}$ and others. According to Eq.(1), the octonion field strength may directly impact the term, $( \partial_0 \textbf{W}_{20}^+ + \nabla \cdot \textbf{W}_2^{i+} )$, which is related to the magnetic vortex clusters. In other words, the magnetic vortex clusters can be appropriately modified and shifted by controlling the field strength or second-torque.

\begin{table*}
\caption{Some arrangement patterns of the magnetic vortices and spin textures within the chiral magnets and frustration magnets, according to the second-torque continuity equation in the 2-quaternion space $\mathbb{H}_e$ .}
\label{tab5}%
\begin{ruledtabular}
\begin{tabular}{@{}lll@{}}
arrangement patterns                                   &    two independent second-torques                                &    examples                      \\
\hline
symmetrical magnetic vortices                          &    identical amplitudes, frequencies, and phase values           &    Refs.[7] and [8]              \\
alternative magnetic vortices                          &    identical amplitudes and frequencies                          &    Refs.[3] and [4]              \\
biased magnetic vortices                               &    identical frequencies                                         &    Refs.[27] and [28]            \\
arbitrary magnetic vortices                            &    arbitrary amplitudes, frequencies, and phase values           &    (many)                        \\
\end{tabular}
\end{ruledtabular}
\end{table*}

\section{Experiment proposal}

According to the second-torque continuity equation, the divergence $\textbf{W}_{20}^+$ of magnetic moments and the second-torque $\textbf{W}_2^{i+}$ are no longer independent, but closely related to each other. On the basis of existing experiments, it is able to suggest a few experiment proposals for several inferences derived from the second-torque continuity equation.

\subsection{Second-torque}

The second-torque is capable of transforming the rotational velocities of some crystal defects (or dopant and others), surrounding by the rotating electric currents within the chiral magnets and frustration magnets. In the 2-quaternion space $\mathbb{H}_e$ , it is easy to find that the second-torque $\textbf{W}_2^{i+}$ will be influenced by the external influence factors (see Ref.[29]), including the magnetic moment $\textbf{L}_2^+$ , electric moment $\textbf{L}_2^{i+}$ , electric field intensity $\textbf{E}^+$ , and magnetic flux density $\textbf{B}$ and so forth. Consequently, making use of the variations of external influence factors, one can vary the second-torque $\textbf{W}_2^{i+}$ , shifting the magnitude and frequency of second-torques.

Making use of the variations of second-torques, one can modify the direction, frequency, and magnitude of the rotational velocities of some crystal defects, within the chiral magnets and frustration magnets and so forth in possession of the rotating electric currents.

\subsection{Unidirectional second-torque}

In the existing experiments for the electric currents flowing through a few crystal defects (or dopant and others), the variations of the external electromagnetic strength and others are able to shift the directions of electric currents. By means of the physical properties of electric currents, the second-torques of crystal defects can be transmitted to the electric currents. The second-torque that the electric currents are subjected to is unidirectional.

Based on the existing experiments for the rotating electric currents, it is able to put forward a suggested experiment, relevant to the unidirectional second-torque. According to the second-torque continuity equation, the second-torque that the rotating electric currents are subjected to is $\textbf{W}_2^{i+}(\omega_0)$, and it is able to transform the divergence of magnetic moments in the rotating electric currents. On the contrary, if one changes the divergence of magnetic moments of the rotating electric currents, the second-torque will be induced correspondingly.

For instance, in a given and arbitrary stream-tube with the varying cross section, in case the magnetic moment, $\textbf{L}_2^+$, of rotating electric currents varies with the fluid-flow or time, the rotating electric currents will be affected by the second-torque, $\textbf{W}_2^{i+}$ . As a result, the rotating electric currents must exert a reverse torque, $-\textbf{W}_2^{i+}$, on the outer wall of the stream-tubes.

By virtue of the measurement of the reverse second-torque, $-\textbf{W}_2^{i+}$ , we can determine the influence of the second-torque, $\textbf{W}_2^{i+}$, on the rotating electric currents. It states that it is able to measure the divergence, $\nabla \cdot \textbf{L}_2^+$, of angular momenta and second-torque $\textbf{W}_2^{i+}$ , in the second-torque continuity equation.

\subsection{Reciprocating second-torque}

In the 2-quaternion space $\mathbb{H}_e$ , there are two types of second-torques, on some crystal defects and others. The first is the unidirectional second-torque, $\textbf{W}_2^{i+} (\omega_1)$, with $\omega_1$ being the angular velocity of rotation in the 2-quaternion space $\mathbb{H}_e$. The second is the reciprocating second-torque, $\textbf{W}_2^{i+} (\omega_2)$ , with $\omega_2$ being the angular velocity of rotation in the 2-quaternion space $\mathbb{H}_e$. Subsequently, the resultant torques, $\{f_{21} \textbf{W}_2^{i+} (\omega_1) + f_{22} \textbf{W}_2^{i+} (\omega_2)\}$, will exert an impact on the electric currents flowing through some crystal defects. Herein $f_{21}$ and $f_{22}$ are the coefficients.

1) When $\omega_1 \neq 0$ and $\omega_2 \neq 0$, it is able to obtain the general and complicated stream patterns of rotating electric currents.

2) When $\omega_1 \neq 0$ and $\omega_2 = 0$, one can achieve the laminar patterns, generated by the electric currents flowing through some crystal defects.

3) When $\omega_1 = 0$ and $\omega_2 \neq 0$, we may attain the turbulent patterns and magnetic vortices, induced by the rotating electric currents flowing through
some crystal defects, especially the magnetic vortex clusters. By virtue of the variation of frequency, $\omega_2$ , of the reciprocating second-torque $\textbf{W}_2^{i+} (\omega_2)$ , it is capable of locally transferring the magnetic vortices, in particular the magnetic vortex clusters.

When the flow of electric currents is affected by the second-torques, the divergence of magnetic moments will change in the magnetic vortex clusters accordingly. On the other hand, the periodic variations of second-torques will result in the periodic variations of the divergence of magnetic moments within the rotating electric currents, inducing the emergence of magnetic vortex clusters.

In the octonion space $\mathbb{O}$ , according to the torque continuity equation, the external torque is able to shift the divergence of angular momenta in the ordinary fluids, transferring the arbitrary vortices and Karman vortex streets of ordinary fluids in the quaternion space $\mathbb{H}_g$. Similarly, according to the second-torque continuity equation, the second-torque is capable of altering the divergence of magnetic moments, transforming the arbitrary magnetic vortices and magnetic vortex clusters of rotating electric currents in the 2-quaternion space $\mathbb{H}_e$ .

As we all know, in terms of some types of charged particles, their magnetic moment and angular momentum may be linearly proportional in a certain range. It means that the second-torques can transform the magnetic vortices and divergence of magnetic moments, for the rotating electric currents in the 2-quaternion space $\mathbb{H}_e$ . Meanwhile the vortices and divergence of angular momenta, for the ordinary fluids in the quaternion space $\mathbb{H}_g$ , will be modified accordingly to a certain extent, due to the linear relationship between magnetic moment and angular momentum. In case the second-torques and other influence factors exert an influence on the rotating electric currents, it will simultaneously lead to the variations of the magnetic vortices, divergence of magnetic moments, vortices, and divergence of angular momenta and others.

These suggested experiments are helpful to further understand the physical properties of magnetic vortex clusters, within the chiral magnets and frustration magnets and others. The physical properties of magnetic vortex clusters are expected to attempt to capture, transfer, and accelerate the elements of rotating electric currents. It may provide a few new technologies for the further developments of some disciplines, including the material science and energy science and others.

\section{Conclusions and discussions}

The quaternion space $\mathbb{H}_g$ is propitious to describe the physical properties of gravitational fields, while the 2-quaternion space $\mathbb{H}_e$ is fit for depicting those of electromagnetic fields. Two quaternion spaces, $\mathbb{H}_g$ and $\mathbb{H}_e$ , are independent of each other. Therefore these two quaternion spaces, which are perpendicular to each other, can be combined together to become one octonion space $\mathbb{O}$. In other words, the octonion space $\mathbb{O}$ is capable of exploring the physical properties of gravitational and electromagnetic fields simultaneously, including the octonion field potential, field strength, field source, linear momentum, angular momentum, torque, and force and so forth.

In the orthogonal coordinate system of the octonion space $\mathbb{O}$ , eight components of one octonion physical quantity are perpendicular to each other, such as the octonion force. Meanwhile the imaginary part and real part of one octonion physical quantity are independent of each other. The octonion force may be equal to zero under some circumstances. Therefore it is able to achieve eight independent equations from the octonion composite force equation, $\mathbb{N}^+ = 0$. In the gravitational and electromagnetic fields, these eight independent equations are capable of deducing eight equilibrium and continuity equations, including the fluid continuity equation, current continuity equation, force equilibrium equation, as well as the second-torque continuity equation discussed in the paper.

The second-torque continuity equation can be applied to explore some physical phenomena relevant to the magnetic vortices of rotating electric currents, within chiral magnets and frustration magnets and others. The gravitational strength, electromagnetic strength, torque, and curl of magnetic moments and others will make a contribution to the second-torque and divergence of magnetic moments, according to the second-torque continuity equation. In the electromagnetic fields, the external second-torque is able to exert an impact on some crystal defects and dopant and so forth within the chiral magnets and frustration magnets and others. It will result in the rotation of the crystal defects and so on in the 2-quaternion space $\mathbb{H}_e$ , altering the magnetic moments. For some types of charged particles, the magnetic moment is proportional to the angular momentum to a certain extent. The emergence of external field strength will inevitably transform the directions of electric currents, and even the arrangement patterns of magnetic vortex clusters within the rotating electric currents.

Obviously, the second-torque continuity equation, $\textbf{N}_{20}^{i+} = 0$, is different from the torque continuity equation, $N_{10}^{i+} = 0$. a) Quaternion spaces. The second-torque continuity equation locates in the 2-quaternion space $\mathbb{H}_e$ , while the torque continuity equation situates in the quaternion space $\mathbb{H}_g$ . b) Fundamental fields. The second-torque continuity equation is propitious to describe the physical properties of electromagnetic fields. And the torque continuity equation is fit for depicting the physical properties of gravitational fields. c) Physical quantities. The second-torque continuity equation explores the relationships among the second-torque and divergence of magnetic moments and so on. And the torque continuity equation researches the relationships among the external torque and divergence of angular momenta and so forth. For some charged particles, they must meet the requirement of the second-torque continuity equation and torque continuity equation simultaneously.

In the paper, the second-torque continuity equation is able to explore a few new physical properties relevant to the flows of rotating electric currents, including the elements of rotating electric currents and second-torque. In the future, it is necessary to consider the second-torque continuity equation, besides the existing equations of hydrodynamics and electrodynamics, when we investigate the flow phenomena of rotating electric currents. It is expected that the second-torque continuity equation will promote the further development of the chiral magnets and frustration magnets and others to some extent.

It is noteworthy that the paper discusses merely a few simple cases, which are associated with the interrelations among the second-torque and the divergence of magnetic moments and others in the rotating electric currents. But it is distinctly revealed that these physical quantities are able to make a contribution to the second-torque continuity equation. The second-torque can alter the magnetic moments and magnetic vortex clusters of rotating electric currents. On the other hand, the divergence of magnetic moments and second-torque obey to the second-torque continuity equation. It is similar to that the mass and linear momentum conform to the fluid continuity equation. In the general case, the motion modes of electric currents will satisfy the demands of the second-torque continuity equation and fluid continuity equation simultaneously. In the future study, we shall research the influence of gravitational strength and electromagnetic strength and so forth on the second-torque continuity equation in the chiral magnets and frustration magnets and others. This will help us to further understand the motion properties of the magnetic vortex clusters in the rotating electric currents.

\begin{acknowledgements}
The author is indebted to the anonymous referees for their valuable comments on the previous manuscripts. This project was supported partially by the National Natural Science Foundation of China under grant number 60677039.
\end{acknowledgements}

\section*{Conflict of interest}
The author has no conflicts to disclose.

\section*{Data Availability}
The data that support the finding of this study are available within this article.

{}

\end{document}